\documentclass[aps,twocolumn]{revtex4}
\usepackage{graphicx}

\begin{document}
\newtheorem{theorem}{Theorem}
\newtheorem{acknowledgement}[theorem]{Acknowledgement}
\newtheorem{algorithm}[theorem]{Algorithm}
\newtheorem{axiom}[theorem]{Axiom}
\newtheorem{claim}[theorem]{Claim}
\newtheorem{conclusion}[theorem]{Conclusion}
\newtheorem{condition}[theorem]{Condition}
\newtheorem{conjecture}[theorem]{Conjecture}
\newtheorem{corollary}[theorem]{Corollary}
\newtheorem{criterion}[theorem]{Criterion}
\newtheorem{definition}[theorem]{Definition}
\newtheorem{example}[theorem]{Example}
\newtheorem{exercise}[theorem]{Exercise}
\newtheorem{lemma}[theorem]{Lemma}
\newtheorem{notation}[theorem]{Notation}
\newtheorem{problem}[theorem]{Problem}
\newtheorem{proposition}[theorem]{Proposition}
\newtheorem{remark}[theorem]{Remark}
\newtheorem{solution}[theorem]{Solution}
\newtheorem{summary}[theorem]{Summary}    
\def\r{{\bf{r}}}
\def\i{{\bf{i}}}
\def\j{{\bf{j}}}
\def\m{{\bf{m}}}
\def\k{{\bf{k}}}
\def\kt{{\tilde{\k}}}
\def\mt{{\hat{t}}}
\def\mG{{\hat{G}}}
\def\mg{{\hat{g}}}
\def\mGa{{\hat{\Gamma}}}
\def\mS{{\hat{\Sigma}}}
\def\mT{{\hat{T}}}
\def\K{{\bf{K}}}
\def\P{{\bf{P}}}
\def\q{{\bf{q}}}
\def\Q{{\bf{Q}}}
\def\p{{\bf{p}}}
\def\x{{\bf{x}}}
\def\X{{\bf{X}}}
\def\Y{{\bf{Y}}}
\def\F{{\bf{F}}}
\def\G{{\bf{G}}}
\def\bG{{\bar{G}}}
\def\mbG{{\hat{\bar{G}}}}
\def\M{{\bf{M}}}
\def\V{\cal V}
\def\tchi{\tilde{\chi}}
\def\tx{\tilde{\bf{x}}}
\def\tk{\tilde{\bf{k}}}
\def\tK{\tilde{\bf{K}}}
\def\tq{\tilde{\bf{q}}}
\def\tQ{\tilde{\bf{Q}}}
\def\si{\sigma}
\def\ep{\epsilon}
\def\hep{{\hat{\epsilon}}}
\def\al{\alpha}
\def\be{\beta}
\def\ep{\epsilon}
\def\bep{\bar{\epsilon}_\K}
\def\mep{\hat{\epsilon}}
\def\up{\uparrow}
\def\de{\delta}
\def\De{\Delta}
\def\up{\uparrow}
\def\dwn{\downarrow}
\def\ksi{\xi}
\def\etha{\eta}
\def\product{\prod}
\def\goto{\rightarrow}
\def\switch{\leftrightarrow}
                           
\title{Comparison of two quantum-cluster approximations}
\author{
Th.\ A.\ Maier
and 
M.\ Jarrell 
}
\address{University of Cincinnati, Cincinnati OH 45221, USA}

\begin{abstract}
We provide microscopic diagrammatic derivations of the 
the Molecular Coherent Potential Approximation (MCA) and
Dynamical Cluster Approximation (DCA) and show that both 
are $\Phi$-derivable.  The MCA (DCA) maps the lattice onto a
self-consistently embedded cluster with open (periodic) 
boundary conditions, and therefore violates (preserves) the
translational symmetry of the original lattice.  As a consequence
of the boundary conditions, the MCA (DCA) converges slowly (quickly) 
with corrections ${\cal{O}}(1/L_c)$ (${\cal{O}}(1/L_c^2)$), where 
$L_c$ is the linear size of the cluster. These analytical results 
are demonstrated numerically for the one-dimensional symmetric 
Falicov-Kimball model. 
\end{abstract}

\maketitle    

\paragraph*{Introduction}
	One of the most active areas in condensed matter physics
is the search for new methods to treat disordered and correlated 
systems.  In these systems, especially in three dimensions or higher, 
approximations which neglect long ranged correlations are generally 
thought to provide a reasonable first approximation for many properties.

Perhaps the most successful of these methods are the Coherent Potential
Approximation (CPA)\cite{CPA} and the Dynamical Mean Field Approximation 
(DMFA)\cite{DMFA_Metzner,DMFA_MullerHartmann,DMFA_Pruschke,DMFA_Georges},
for disordered and correlated systems, respectively.  Although these
approximations have different origins, they share a common 
microscopic definition.  Both the DMFA\cite{DMFA_MullerHartmann} and the 
CPA\cite{DCA_Jarrell1} may be defined as theories which completely 
neglect momentum conservation at all internal diagrammatic vertices.  
When this principle is applied, the diagrammatic expansion for the 
irreducible quantities in each approximation collapse onto that of a 
self-consistently embedded impurity problem.  

Many researchers have actively searched for a technique to restore 
non-local corrections to these approaches.  Here, we discuss just
two approaches which are fully causal and self-consistent: the 
Molecular Coherent Potential Approximation (MCA)\cite{MCPA,agonis} 
and the Dynamical Cluster Approximation 
(DCA)\cite{DCA_Hettler1,DCA_Hettler2,DCA_Maier1,DCA_Jarrell1}. 
Recently the Cellular Dynamical Mean Field 
Approach\cite{kotliar} was proposed for ordered correlated systems, 
while the Molecular Coherent Potential Approximation  has traditionally 
been applied to disordered systems. Since both methods share a common 
microscopic definition we use the term MCA to refer to both techniques 
in the following.

While the MCA is traditionally defined in the real space of the 
lattice, the DCA is traditionally defined in its reciprocal space.  
In the MCA, the system lattice is split into a series of identical 
molecules.  Interactions between the molecules are treated in a 
mean-field approximation, while interactions within the molecule 
are explicitly accounted for.    In the DCA, the reciprocal space 
of the lattice is split into cells, and momentum conservation is 
neglected for momentum transfers within each cell while it is 
(partially) conserved for transfers between the cells.  These 
approximations share many features in common: they both map the 
lattice problem onto that of a self-consistently embedded cluster 
problem.  Both recover the single site approximation (CPA or DMFA) 
when the cluster size reduces to one and become exact as the cluster
size diverges.  Both are fully causal\cite{MCPA,DCA_Hettler2}, and 
provided that the clusters are chosen correctly\cite{DCA_Jarrell1}, 
they maintain the point group symmetry of the original lattice 
problem.  Here, we provide a microscopic diagrammatic derivation 
of both the MCA and the DCA, and explore their convergence
with increasing cluster size.

\paragraph*{Formalism}
For simple Hubbard-like models, momentum conservation 
at each vertex is completely described by the Laue function
\begin{equation}
\label{eq:Laue}
\De=\sum_\x e^{i\x\cdot(\k_1+\k_2-\k_3-\k_4)}=N\de_{\k_1+\k_2,\k_3+\k_4}\,,
\end{equation}
where $\k_1$, $\k_2$ ($\k_3$, $\k_4$) are the momenta entering (leaving) 
the vertex. M\"uller-Hartmann \cite{DMFA_MullerHartmann}
showed that the Dynamical Mean Field (DMF) theory may be derived by 
completely ignoring momentum conservation at each internal vertex by 
setting $\De=1$.  Then, one may freely sum over all 
of the internal momentum labels, and the graphs for the generating functional 
$\Phi$ and its irreducible derivatives,  contain only local propagators.

The DCA and MCA techniques may also be defined by their respective Laue 
functions.  Since our object is to define  cluster methods we divide the 
original lattice of $N$ sites into $N/N_c$ clusters (molecules), each 
composed of $N_c=L_c^D$ sites, where $D$ is the dimensionality. We use the 
coordinate $\tx$ to label the origin of the clusters and $\X$ to label the 
$N_c$ sites within a cluster, so that the site indices of the original 
lattice $\x=\X+\tx$.  The points $\tx$ form a lattice with a reciprocal 
space labeled by $\tk$.  The reciprocal space corresponding to the sites 
$\X$ within a cluster shall be labeled $\K$, with 
$K_\alpha=n_\alpha \cdot 2\pi/L_c$ and integer 
$n_\alpha$. Then $\k=\K+\tk$. Note that $e^{i\K\cdot \tx}=1$ since a 
component of $\tx$ must take the form $m_\alpha L_c$ with integer $m_\alpha$. 

In the MCA, we approximate the Laue function by
\begin{equation}
\label{eq:LMCA}
\De_{MC}=\sum_{\X}e^{i\X\cdot (\K_1+\K_2-\K_3-\K_4+\tk_1+\tk_2-\tk_3-\tk_4)}\,.
\end{equation}
Thus, the MCA omits the phase factors $e^{i\tk\cdot\tx}$ resulting from the 
position of the cluster in the original lattice but retains the (far less 
important) phase factors $e^{i\tk\cdot\X}$ associated with the position 
within a cluster. In the DCA we also omit the phase factors  $e^{i\tk\cdot\X}$, 
so that
\begin{equation}
\label{eq:LDCA}
\Delta_{DC}=N_c\de_{\K_1+\K_2,\K_3+\K_4}\,.
\end{equation}
Both the MCA and DCA Laue functions recover the exact result when 
$N_c\to\infty$ and the DMFA result, $\Delta=1$, when $N_c=1$.

If we apply the MCA Laue function Eq.~\ref{eq:LMCA} to diagrams in $\Phi$, 
then each Green function leg is replaced by the MCA coarse-grained Green 
function
\begin{eqnarray}
\label{eq:cgMCA}
\bG(\X_1,\X_2;\tx=0)&=&\nonumber\\
&&\hspace*{-3cm}\frac{1}{N^2}\!\!\!\sum_{\stackrel{\K_1,\K_2}{\tk_1,\tk_2}}
e^{i(\K_1+\tk_1)\cdot\X_1}G(\K_1,\K_2;\tk_1,\tk_2)e^{-i(\K_2+\tk_2)\cdot\X_2}
=\nonumber\\
&&\hspace*{-3cm}\frac{N_c^2}{N^2}\sum_{\tk_1,\tk_2}G(\X_1,\X_2,\tk_1,\tk_2)\,,
\end{eqnarray}
or in matrix notation for the cluster sites $\X_1$ and $\X_2$
\begin{equation}
\label{eq:cgGMCA}
\hat{\bG}=\frac{N_c}{N}\sum_{\tk}\mG(\tk)\,.
\end{equation}   
The summations of the cluster sites $\X$ remain to be performed.
Note that the inclusion of the phase factors 
$e^{i\tk\cdot\X}$ in the MCA Laue-function Eq.~\ref{eq:LMCA} leads 
directly to a cluster approach formulated in real space that violates 
translational invariance. Therefore the Green function is 
a function of two cluster momenta $\K_1$, $\K_2$ or two sites 
$\X_1$, $\X_2$ respectively.

If we apply the DCA Laue function Eq.~\ref{eq:LDCA}, Green function legs 
in $\Phi$ are replaced by the DCA coarse grained Green function
\begin{equation}
\label{eq:cgGDCA}
\bG(\K)=\frac{N_c}{N}\sum_{\tk}G(\K,\tk)\,,
\end{equation} 
since Green functions can be freely summed over the $\tk$ vectors within 
a cell about the cluster momentum $\K$.  (We have dropped the frequency
dependence for notational convenience.)  As a result, $\Phi$ is a 
functional of the coarse grained Green function $\bar{G}(\K)$ and thus 
depends on the cluster momenta $\K$ only. 

To establish a connection between the cluster and the lattice we minimize 
the lattice free energy
\begin{equation}
F= -k_B T\left(
\Phi_{c}-\mbox{tr}\left[{\bf{\Sigma}} {\bf{G}}\right] 
+\mbox{tr}\ln\left[{\bf{G}}\right]\right)
\label{F_CA}
\end{equation}
where $\Phi_{c}$ is the generating functional calculated with the
coarse-grained propagators, ${\bf{\Sigma}}$ is the lattice self-energy 
and ${\bf{G}}$ is the full lattice Green function.  The trace 
indicates summation over frequency, momentum and 
spin.  As we have discussed elsewhere, only the compact part of the 
free energy, $\Phi$, is coarse-grained.  $F$ is stationary with respect 
to ${\bf{G}}$ when $\frac{\delta F}{\delta G}=0$.
This happens for the MCA if we estimate the lattice self energy as
\begin{eqnarray}
\label{eq:MDMFS}
\Sigma(\K_1,\K_2;\tk_1,\tk_2)&=&\nonumber\\
&&\hspace*{-3cm}\sum_{\X_1,\X_2} e^{-i(\K_1+\tk_1)\cdot\X_1}
\Sigma_{MC}(\X_1,\X_2) e^{i(\K_2+\tk_2)\cdot\X_2}\,.
\end{eqnarray}
Thus, the corresponding lattice single-particle propagator reads in matrix 
notation 
\begin{equation}
\label{eq:GXk}
\mG(\tk,z)=\left[zI-\mep(\tk)-\mS_{MC}(z)\right]^{-1}\,,
\end{equation}
where the dispersion $\mep(\tk)$ and self-energy $\mS_{MC}(z)$ are 
matrices in cluster real space with 
\begin{eqnarray}
[\mep(\tk)]_{\X_1\X_2}&=&\epsilon(\X_1-\X_2,\tk)\\\nonumber
&=&\frac{1}{N_c}\sum_\K e^{i(\K+\tk)(\X_1-\X_2)}\epsilon_{\K+\kt}
\end{eqnarray}
being the intracluster Fourier transform of the dispersion.
For the DCA, $\Sigma(\k)=\Sigma_{DC}(\K)$ is the proper 
approximation for the lattice self energy corresponding to $\Phi_{DC}$.
The corresponding lattice single-particle propagator is then given by 
\begin{equation}
G(\K,\tk;z) =\frac{1}{z-\ep_{\K+\tk}-\Sigma_{DC}(\K,z) } \,.
\label{G_DCA}
\end{equation}
Both the MCA and DCA are optimized when we equate the lattice and
cluster self energies.  A similar relation holds for two-particle
quantities.  Thus, with few exceptions\cite{exception}, only the 
irreducible quantities on the cluster and lattice correspond one-to-one.

The MCA (DCA) algorithm, with steps {\bf{A}}$\to${\bf{D}}, follows directly: 
{\bf{A}} we first make an initial guess for the cluster self-energy matrix 
$\Sigma$.  {\bf{B}} This is used 
with Eqs.~\ref{eq:cgGMCA} and \ref{eq:GXk} (\ref{eq:cgGDCA} and \ref{G_DCA}) 
to calculate the coarse-grained Green function $\bar G$.  {\bf{C}} The 
cluster excluded Green function $\hat{\cal{G}}=[\mbG^{-1}+\mS_{MC}]^{-1}$
(${\cal{G}}(\K)= [\bG(\K)^{-1}+\Sigma_{DC}(\K)]^{-1}$)
is defined to avoid overcounting self energy corrections on the cluster.  
It is used to {\bf{D}} compute a new estimate for the cluster self-energy 
which is used to reinitialize the process.  Once convergence is reached, 
the irreducible quantities on the cluster may be used to calculate the 
corresponding lattice quantities.

In order to compare the character of the two different cluster approaches 
as a function of the cluster size $N_c$ it is instructive to rewrite the 
corresponding coarse grained Green-functions Eqs.~\ref{eq:cgGMCA} and 
\ref{eq:cgGDCA} to suitable forms by making use of the independence 
of the self-energy $\Sigma$ on the integration variable $\tk$. For the 
MCA coarse grained Green function we find
\begin{equation}
\label{eq:hMCA}
\mbG(z)=\left[zI-\hep_o-\mS_{MC}(z)-\mGa_{MC}(z)\right]^{-1}\,,
\end{equation}
with the ``cluster-local'' energy $\hep_o=N_c/N\sum_{\tk}\hat{\epsilon}(\tk)$. 
For the DCA we 
obtain a similar expression
\begin{equation}
\label{eq:hDCA}
\bG(\K,z)=\left[z-\bep-\Sigma_{DC}(\K,z)-\Gamma_{DC}(\K,z)\right]^{-1}\,,
\end{equation}
with the coarse grained average $\bep=N_c/N\sum_{\tk}\ep(\K,\tk)$.
The hybridization functions $\mGa_{MC/DC}(z)$ describe the coupling of the 
cluster to the mean field representing the remainder of the system.

The behavior of $\Gamma$ for large $N_c$ is important. For the MCA, $\Gamma$
averaged over the cluster and frequency 
\begin{equation}
\label{eq:GbMCA}
\bar{\Gamma}_{MC}=
\frac{1}{N_c}\sum_{\X_1,\X_2}\Gamma_{MC}(\X_1,\X_2)
\sim{\cal O}\left(\frac{2D}{L_c}\right)\,, 
\end{equation}
where $L_c=N_c^{1/D}$ is the linear cluster size.  A detailed derivation
of this form will be published elsewhere.  However, since in the MCA the 
cluster is defined in real space with open boundary conditions, this form 
is evident since only the sites on the surface $\propto 2D\cdot L_c^{D-1}$ 
of the cluster couple to the effective medium.  For the DCA we have previously 
shown that $\Gamma(\K)\sim {\cal O}(1/N_c^{2/D})$ \cite{DCA_Maier1} 
so that we obtain for the average hybridization of the DCA cluster to the 
effective medium 
\begin{equation}
\label{eq:GavDC}
\bar{\Gamma}_{DC}=
\frac{1}{N_c}\sum_{\K}\Gamma_{DC}(\K)\sim{\cal O}\left(\frac{1}{L_c^2}\right)\,.
\end{equation}
The DCA coarse graining results in a cluster in $\K$-space; thus, the 
corresponding real space cluster has periodic boundary conditions, and 
each site in the cluster has the same hybridization strength $\bar\Gamma$ 
with the host.

In both the MCA and the DCA, the average hybridization strength acts
as the small parameter.  The approximation performed by the 
MCA (DCA) is to replace the lattice Green function 
$\mG(\tk)=[zI-\hep(\tk)-\mS(\tk,z)]^{-1}$ 
($G(\K,\tk,z)=[z-\ep_{\K+\tk}-\Sigma(\K,\tk,z)]^{-1}$) by its 
coarse grained quantity $\mbG$ ($\bG(\K)$) in diagrams for the self-energy 
$\Sigma$.  Once the sums over $\tk$ are performed, all terms which are lower 
order in $1/L_c$ than $\Gamma$ vanish.  Thus the MCA (DCA) is an approximation 
with corrections of order $\bar{\Gamma}$ $\sim{\cal O}(1/L_c)$
($\sim{\cal O}(1/L_c^2)$).

\paragraph*{Numerical Results}
To illustrate the differences in convergence with cluster size $N_c$
we performed MCA and DCA simulations for the symmetric one-dimensional 
(1D) Falicov-Kimball model (FKM). At half filling the FKM Hamiltonian reads
\begin{equation}
H=-t\sum_i (d_i^\dagger d_{i+1}^{}+h.c.)+U\sum_i(n_i^d-1/2)(n_i^f-1/2)\,,
\end{equation} 
with the number operators $n_i^d=d_i^\dagger d_i^{}$ and 
$n_i^f=f_i^\dagger f_i^{}$ and the Coulomb repulsion $U$ between $d$ 
and $f$ electrons residing on the same site.  The FKM can be considered as 
a simplified Hubbard model with only one spin-species ($d$) being allowed 
to hop. However it still shows a complex phase diagram including a Mott 
transition and Ising-like charge ordering with the corresponding 
transition temperature $T_c$ being zero in 1D. The dispersion in (1D) 
$\epsilon_k=2t\cos k$; thus for $t=1/4$ the bandwidth $W=1$ which we use 
as unit of energy. To simulate the effective cluster models of the MCA 
an the DCA we use a quantum Monte Carlo (QMC) approach described 
in\cite{DCA_Hettler2}. 

\begin{figure}[htb]
\includegraphics*[width=3.0in]{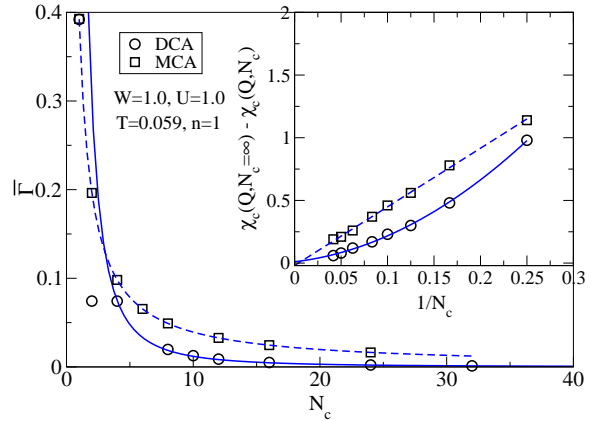}
\caption{The average integrated hybridization strengths $\bar{\Gamma}$ of 
the MCA (squares) and DCA (circles) versus the cluster size $N_c$ when 
$\beta=17$ and $U=W=1$. The solid and dashed lines represent the fits 
$1.1946/N_c^2$ and $0.3361/N_c$ respectively. Inset: Convergence of the 
cluster charge susceptibility for $Q=\pi$. The solid and dashed lines 
are quadratic and linear fits, respectively.} 
\label{fig:GvsL}
\end{figure}

To check the scaling relations Eqs.~\ref{eq:GbMCA} and \ref{eq:GavDC}, 
we show in Fig.\ref{fig:GvsL} the average hybridization functions 
$\bar{\Gamma}_{MC}$ and $\bar{\Gamma}_{DC}$ for the MCA and DCA respectively 
at the inverse temperature $\beta=17$ for $U=W=1$.   
For $N_c=1$ both approaches are equivalent to the DMFA and thus 
$\bar{\Gamma}_{MC}=\bar{\Gamma}_{DC}$. For increasing $N_c$ 
$\bar{\Gamma}_{MC}$ can be fitted by $0.3361/N_c$ and $\bar{\Gamma}_{DC}$ 
by $1.1946/N_c^2$ when $N_c>2$. 
Cluster quantities, such as the self energy and cluster susceptibilities, 
are expected to converge with increasing $N_c$ like $\bar{\Gamma}$.  This 
is illustrated in the inset for the staggered ($Q=\pi$) charge 
susceptibility $\chi_c({\bf Q})$ of the cluster.

Since only the compact parts represented by $\Phi$ of the lattice free energy 
(Eq.~\ref{F_CA}) are coarse-grained, this scaling is expected to break down 
when lattice quantities, such as the lattice charge susceptibility, are 
calculated.  The susceptibility of the cluster $\chi_c(Q)$ cannot 
diverge for any finite $N_c$; whereas the lattice $\chi(Q)$ diverges 
at the transition temperature $T_c$ to the charge ordered phase.  Note that 
the residual mean field character of both methods can result in finite 
transition temperatures $T_c>0$ for finite $N_c<\infty$. However as $N_c$ 
increases, this residual mean field character decreases gradually and thus 
increased fluctuations should drive the solution to the exact result $T_c=0$.

\begin{figure}
\includegraphics*[width=3.0in]{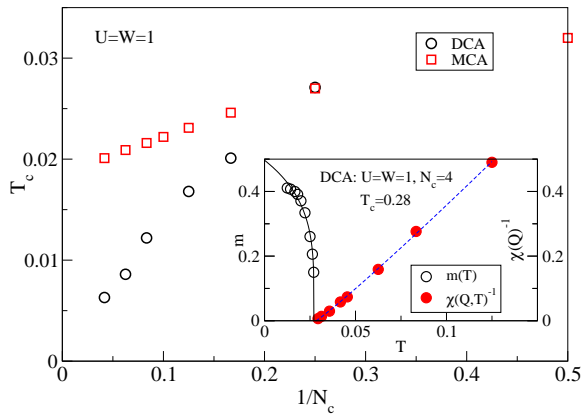}
\caption{The transition temperature $T_c$ for the DCA (circles) and 
MCA (squares) when $U=W=1$ versus the cluster size $N_c$. For all 
values of $N_c$ the DCA prediction is closer to the exact result 
($T_c=0$). Inset: Order parameter $m(T)$ and inverse charge 
susceptibility $\chi(Q)^{-1}$ versus temperature. The solid (dashed) line 
represents a  fit to the functions $m(T)\propto (T_c-T)^\beta$ with 
$\beta=0.245$ ($\chi(T)\propto (T-T_c)^{-\gamma}$ with $\gamma=1.07$).} 
\label{fig:phase}
\end{figure}

In the DCA\cite{DCA_Hettler2}, $\chi(Q)$ is calculated by first extracting 
the corresponding vertex function from the cluster simulation.  This is then 
used in a Bethe-Salpeter equation to calculate $\chi(Q)$.  $T_c$ is 
calculated by extrapolating $\chi(Q)^{-1}$ to zero using the function 
$\chi(Q)^{-1}\propto (T-T_c)^\gamma$ (see inset to Fig.\ref{fig:phase}).
This procedure is difficult, if not impossible, in the MCA due to the lack 
of translational invariance. Here, we calculate the the order parameter 
$m(T)=1/N_c\sum_i(-1)^i\langle n_i^d\rangle$ in the symmetry broken phase.  
$T_c$ is then obtained from extrapolating $m(T)$ to zero using the function 
$m(T)\propto (T_c-T)^\beta$. For the DCA this extrapolation is shown by the 
solid line in the inset to Fig.\ref{fig:phase} for $N_c=4$.  The values for 
$T_c$ obtained from the calculation in the symmetry broken phase and in the 
unbroken phase must agree, since as we have shown above, both the DCA 
and MCA are $\Phi$-derivable.  This is illustrated in Fig.\ref{fig:phase} for 
the DCA.    

A comparison of the DCA and MCA estimate of $T_c$ is presented in
Fig.~\ref{fig:phase}.  $T_c$ obtained from MCA (squares) is larger than 
$T_c$ obtained from DCA (circles). Moreover we find that the DCA result 
seems to scale to zero almost linearly in $1/N_c$ (for large enough $N_c$), 
whereas the MCA does not show any scaling form and in fact seems to tend to 
a finite value for $T_c$ as $N_c\rightarrow\infty$. This striking 
difference of the two methods can be attributed to the different 
boundary conditions. The open boundary conditions of the MCA cluster result 
in a large surface contribution so that $\bar{\Gamma}_{MC}>\bar{\Gamma}_{DC}$.  
This engenders pronounced mean field behavior that stabilizes the finite 
temperature transition for the cluster sizes treated here. For larger 
clusters we expect the bulk contribution to the MCA free energy to dominate 
so that $T_c$ should fall to zero. 

Complementary results are found in simulations of {\em{finite-sized}} 
systems.  In general, systems with open boundary conditions are expected 
to have a surface contribution in the free energy of order 
${\cal{O}}(1/L_c)$\cite{M_Fisher_72}.   This term is absent in systems 
with periodic boundary conditions. As a result, simulations of 
finite-sized systems with periodic boundary conditions converge much 
more quickly than those with open boundary conditions\cite{D_Landau_76}.     

\paragraph*{Summary.}  By defining appropriate Laue functions, we
provide microscopic diagrammatic derivations of the MCA and DCA.   We 
show that they are $\Phi$-derivable, and that the lattice free energy
is optimized by equating the irreducible quantities on the lattice to 
those on the cluster.  The MCA maps the lattice to a cluster with open 
boundaries and consequently, the cluster violates translational 
invariance.  In contrast, the DCA cluster has periodic boundary 
conditions, and therefore preserves the translational invariance of 
the lattice. This difference in the boundary conditions translates 
directly to different asymptotic behaviors for large clusters $N_c$. 
As we find analytically as well as numerically, the surface 
contributions in the MCA lead to an average hybridization 
$\bar\Gamma$  of the cluster  to the mean field that scales like 
$1/L_c$ as compared to the $1/L_c^2$ scaling of the DCA.   Since 
$\bar\Gamma$ acts as the small parameter for these approximation 
schemes, the DCA converges much more quickly than the MCA.  These 
effects are more pronounced near a transition, where the large 
surface contribution of the MCA stabilizes the mean-field character
of the transition.  Consequently, the DCA result for the transition 
temperature $T_c$ of the 1D symmetric FKM model scales almost like 
$1/N_c$ to the exact result $T_c=0$, whereas the MCA result converges 
very slowly. Since the origin of this difference lies in the different 
boundary conditions we expect this primacy of the DCA over the MCA 
to hold generally for any model of electrons moving on a lattice.      

\paragraph*{Acknowledgements} We acknowledge useful conversations with 
N.\ Bl\"umer,
A.\ Gonis,
M.\ Hettler,
H.R.\ Krishnamurthy,
D.P.\ Landau,
Th.\ Pruschke,
Th.\ Schulthess,
W.\ Shelton,
and 
A.\ Voigt. 
This work was supported by NSF grant DMR-0073308.

\end{document}